\documentclass[aps,prl,reprint, groupedaddress]{revtex4-2}
\usepackage{graphicx} 
\usepackage{amsmath}
\usepackage{amsfonts}
\usepackage{amssymb}
\usepackage{tikz}
\usetikzlibrary{shapes.geometric}
\usepackage{booktabs}
\usepackage{longtable}            
\usepackage{rotating}
\usepackage{ifthen}
\usepackage{colortbl}
\usepackage{soul}
\usepackage{makecell}

\definecolor{shivugreen}{rgb}{0.0117647059, 0.5568627451, 0.431372549}

\newcommand{\custommarker}[1]{%
  \tikz[baseline=(base.base), scale=1, transform shape]{
    \node (base) at (0,0) {};  

    \ifthenelse{\equal{#1}{diamond}}{
      \draw[fill=shivugreen, draw=black, opacity=0.5]
        (0,0.14cm) -- (0.1cm,0) -- (0,-0.14cm) -- (-0.1cm,0) -- cycle;
    }{
    \ifthenelse{\equal{#1}{smalldiamond}}{
      \draw[fill=shivugreen, draw=black, opacity=0.5]
        (0,0.09cm) -- (0.06cm,0) -- (0,-0.09cm) -- (-0.06cm,0) -- cycle;
    }{
    \ifthenelse{\equal{#1}{orangediamond}}{
      \draw[fill=orange, draw=black, opacity=0.5]
        (0,0.14cm) -- (0.1cm,0) -- (0,-0.14cm) -- (-0.1cm,0) -- cycle;
    }{

    \ifthenelse{\equal{#1}{circle}}{
      \draw[fill=black, draw=black, opacity=0.5]
        (0,0) circle (0.1cm);
    }{
    \ifthenelse{\equal{#1}{smallcircle}}{
      \draw[fill=black, draw=black, opacity=0.5]
        (0,0) circle (0.05cm);
    }{

    \ifthenelse{\equal{#1}{triangle}}{
      \draw[fill=black, draw=black, opacity=0.5]
        (-0.1cm,-0.07cm) -- (0.1cm,-0.07cm) -- (0,0.13cm) -- cycle;
    }{
    \ifthenelse{\equal{#1}{orangetriangle}}{
      \draw[fill=orange, draw=black, opacity=0.5]
        (-0.1cm,-0.07cm) -- (0.1cm,-0.07cm) -- (0,0.13cm) -- cycle;
    }{

    \ifthenelse{\equal{#1}{invertedblue}}{
      \draw[fill=blue, draw=black, opacity=0.5]
        (0,-0.13cm) -- (-0.1cm,0.07cm) -- (0.1cm,0.07cm) -- cycle;
    }{

    \ifthenelse{\equal{#1}{smallsquare}}{
      \draw[fill=shivugreen, draw=black, opacity=0.5]
        (-0.05cm,-0.05cm) rectangle (0.05cm,0.05cm);
    }{
    \ifthenelse{\equal{#1}{bigsquare}}{
      \draw[fill=shivugreen, draw=black, opacity=0.5]
        (-0.15cm,-0.15cm) rectangle (0.15cm,0.15cm);
    }{
    \ifthenelse{\equal{#1}{orangesquare}}{
      \draw[fill=orange, draw=black, opacity=0.5]
        (-0.1cm,-0.1cm) rectangle (0.1cm,0.1cm);
    }{
    \ifthenelse{\equal{#1}{greensquare}}{
      \draw[fill=shivugreen, draw=black, opacity=0.5]
        (-0.1cm,-0.1cm) rectangle (0.1cm,0.1cm);
    }{

    \ifthenelse{\equal{#1}{blackstar}}{
      \node[star, star points=5, star point ratio=2.25,
            minimum size=0.3cm, inner sep=0pt, outer sep=0pt,
            fill=black, draw=black, opacity=0.5] at (0,0) {};
    }{
    \ifthenelse{\equal{#1}{orangestar}}{
      \node[star, star points=5, star point ratio=2.25,
            minimum size=0.3cm, inner sep=0pt, outer sep=0pt,
            fill=orange, draw=black, opacity=0.5] at (0,0) {};
    }{
    \ifthenelse{\equal{#1}{greenstar}}{
      \node[star, star points=5, star point ratio=2.25,
            minimum size=0.3cm, inner sep=0pt, outer sep=0pt,
            fill=shivugreen, draw=black, opacity=0.5] at (0,0) {};
    }{

    \draw[fill=green, draw=black, opacity=0.5]
      (-0.075cm,-0.075cm) rectangle (0.075cm,0.075cm);

    }}}}}}}}}}}}}}}}
}

\begin{document}

\title{Scale dependence of segregation patterns in the filling of silos}

\author{Shivakumar Athani$^{1}$}
\author{Benjy Marks$^{1}$}%
\author{Fran\c{c}ois Guillard$^{1}$}%
\author{Alistair Gillespie$^{2}$}%
\author{Itai Einav$^{1}$}%

\affiliation{$^1$School of Civil Engineering, The University of Sydney, Sydney 2006 NSW, Australia\\
$^2$Rio Tinto, Brisbane, Queensland, Australia}

\date{\today}
\begin{abstract}
 Size segregation in granular flows is a well-known phenomenon: laboratory experiments consistently show that large particles migrate toward silo walls during filling, while smaller particles concentrate near the center. Paradoxically, field observations in large-scale industrial silos often report the opposite pattern, challenging these findings. We demarcate these patterns through a systematic experimental study spanning a range of dimensionless numbers relevant to bidisperse granular flows in quasi-2D silos under both dry and immersed conditions, varying container geometry and fluid viscosity. Image analysis reveals that the observed patterns are governed by two key dimensionless parameters: the slenderness of the silo and the Stokes number, which encapsulates the balance between particle inertia and viscous drag. Our results demonstrate the role of fluids on segregation dynamics and provide a unified scaling framework that reconciles laboratory- and field-scale observations in air.
\end{abstract}

\maketitle

When granular materials composed of two sizes of grains (termed bidisperse) are poured into a silo, the flowing material tends to segregate by size. In laboratory-scale experiments (typically 10–30 cm wide) conducted in air or vacuum, this causes larger grains to migrate toward the container periphery, near the silo walls, while the center is primarily occupied by smaller grains \cite{hastie2000segregation,fan2014modelling,shimokawa2015pattern,xiao2019continuum}. Additionally, some studies have shown the stratification of alternating layers of small and large particles during segregation \cite{makse1997spontaneous,fan2017segregation}. The precise extent of this size separation has been shown to depend on the feed rate and container geometry \cite{fan2017segregation}, and has been explored using a range of continuum, discrete and cellular automaton models, \textit{e.g.}, \cite{gray1997pattern, fan2017segregation,dissanayake2025modelling}.

By contrast, large-scale observations in industry often show the opposite pattern, with the finer grains rather than the larger ones near the walls, during the filling of dry silos (often 10–50 m wide) \cite{engblom2012segregation}. This behavior has been attributed to interactions between the particles and the surrounding air that circulates within the silo \cite{dyroy2001system,mosby1996segregation,zigan2007air,zigan2008theoretical,schulze2021powders,engblom2012effects,engblom2012segregation}.

Unlike in the laboratory, studying internal dynamics in large-scale silos is challenging given the scale, opacity, and dusty environment. To overcome this, we develop a unified experimental framework based solely on laboratory-scale experiments to elucidate the physics of segregation patterns which emerge depending on the surrounding fluid drag. Our approach combines granular filling of dry and fluid-immersed silos with image analysis to identify three distinct segregation patterns: the two previously described, and a differential settlement pattern --- commonly observed in sedimentation --- where larger heavy grains are below smaller light ones \cite{weiland1984instabilities,abbas2006dynamics,snabre2009size,deboeuf2011segregation,li2022hindered}. By identifying key dimensionless parameters, we demonstrate that a simple phase diagram captures the transitions between these patterns.

Granular flow in silos under immersed conditions has shown that the surrounding fluid can strongly influence particle trajectories. This effect has been observed in both monodisperse systems \cite{koivisto2017sands,fan2022discharge} and bidisperse systems, where increasing fluid viscosity was found to suppress segregation \cite{samadani2000segregation,samadani2001angle}. However, unlike the present study, the experimental protocols in these earlier works considered the introduction of dry particles from a vertically moving tube positioned just above the immersed granular heap throughout the filling process, strongly limiting the falling distance and velocity of the particles. In contrast, our study uses a fully fluid-saturated system, that closely mimics real silo conditions, enabling a systematic investigation of how immersion and fluid properties influence segregation dynamics across regimes.

\begin{figure}
    \centering
    \includegraphics{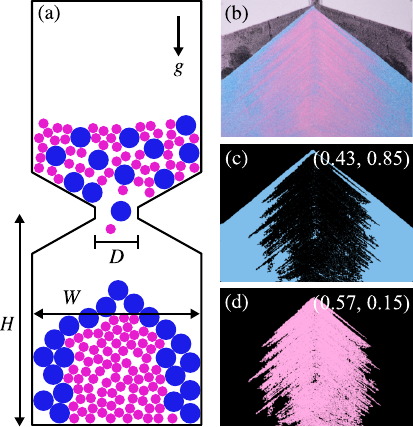}
    \caption{Experimental set-up and image analysis methodology. (a) Schematic representation of the experimental silo, which is composed of two acrylic sheets separated by a distance $T$, with particles of two sizes immersed in a fluid between the sheets. Each chamber has height $H$, width $W$ and gap opening $D$. (b) An example of a segregated heap after filling a silo in air. (c) and (d) segmentation of large and small particles based on their colour, respectively. The values on the respective plots indicate the normalized second moments of area ($\hat I_{xx}^{\alpha}$, $\hat I_{yy}^{\alpha}$), $\hat I_{xx}^{\alpha} =\frac{I_{xx}^{\alpha}}{I_{xx}^{s}+I_{xx}^{b}}$, where $\alpha \in \{s,b \}$ with $s$ and $b$ indicating small and big particles, respectively.}
    \label{fig:lab_model}
\end{figure}

To examine the scale dependence of segregation patterns, we conducted small-scale laboratory experiments on the filling of silos under gravity $g$, with the silo immersed in a fluid of viscosity $\eta$ and density $\rho_f$, as illustrated in Figure~\ref{fig:lab_model}. The granular material consisted of particles of density $\rho_p$, either made of glass ($\rho_p=2500~\mathrm{kg/m^3}$) or Alumina ($\rho_p=3970~\mathrm{kg/m^3}$), sieved into two narrow size fractions with nominal diameters $d_s \pm 20\%$ and $d_b \pm 20\%$, where $d_b \approx 3d_s$. These were mixed at equal gravimetric concentrations, resulting in a volume averaged particle diameter $d = \tfrac{1}{2}(d_s + d_b)$.

To enable visual identification of spatial segregation, particles were color-coded by size; see the Appendix for details. The surrounding fluid medium was varied among four cases:  
air ($\rho_f = 1~\mathrm{kg/m^3}$, $\eta = 10^{-5}~\mathrm{kg/(ms)}$),  
water ($\rho_f = 1{,}000~\mathrm{kg/m^3}$, $\eta = 10^{-3}~\mathrm{kg/(m\,s)}$),  
propanol ($\rho_f = 786~\mathrm{kg/m^3}$, $\eta = 2.2 \times 10^{-3}~\mathrm{kg/(m\,s)}$), and  
sunflower oil ($\rho_f = 940~\mathrm{kg/m^3}$, $\eta = 5 \times 10^{-2}~\mathrm{kg/(m\,s)}$).

The experiments were performed in Hele-Shaw cells of various widths $W$ and orifice widths $D$. The thickness $T$ was 0.2 cm for the experiments with alumina powder, and 1 cm for the ones using glass beads, to ensure flowability. To avoid jamming and maintain reliable flow, most experiments satisfied $W/D \gtrsim 10$ and $D/d > 10$. Prior to each run, the particles were thoroughly mixed and then allowed to flow under gravity through the orifice into a lower chamber.

A summary of the experimental parameters is provided in Table~\ref{tab:all}. As discussed by \cite{zigan2007air}, the problem of silo filling involves a range of dimensionless numbers. In this work, we consider only three non-dimensional numbers. The first is geometrical, and refers to the aspect ratio, or slenderness, of the silo:
\begin{equation}\label{eq:aspect_ratio}
    A_r=\frac{H}{W}.
\end{equation} 

The second non-dimensional parameter, which is often used to analyse fluid-immersed granular system, is the Stokes number, the ratio between the particle response time to the characteristic fluid time $St=\tfrac{\tau_\mathrm{p}}{\tau_\mathrm{f}}$ \citep[e.g.][]{courrech2003granular,bougouin2018granular}. The particle response time $\tau_\mathrm{p}=\tfrac{\rho_pd^2}{18\eta}$ represents the time over which a particle responds to changes in the surrounding fluid flow. The characteristic fluid time can be defined as $\tau_\mathrm{f}=H/v_t$, where $v_t=\tfrac{(\rho_p - \rho_f)gd^2}{18\eta}$ represents the terminal velocity of free falling particles in the viscous fluid
\cite{courrech2003granular,guazzelli2011fluctuations,bougouin2018granular,poddar2024experimental}. In this problem, $H$ is the relevant length scale as it determines whether the particles have sufficient distance to reach their terminal velocity. If the particles' terminal velocity is reached, then the particle trajectories are only influenced by the surrounding fluid flow, otherwise inertia drives their motion. The use of $H$ instead of $d$ in $\tau_f$ follows \cite{engblom2012effects,engblom2012segregation} to capture fluid-driven segregation. The Stokes number for the silo problem is here defined by:
\begin{equation}\label{eq:stk}
St=\frac{\rho_p d^2 v_t}{18\eta H}.
\end{equation}

Finally, we define the Reynolds number of the particles through the ratio of inertial $f_i=\frac{\rho_f v_t^2}{d}$ to viscous $f_v=\frac{\eta v_t}{d^2}$ forces in terms of their terminal velocity \cite{turton1986short,guazzelli2011fluctuations,woodcock2012bulk}: 
\begin{equation}\label{eq:rep}
Re_p=\frac{\rho_f v_t d}{\eta}.
\end{equation}

\begin{table*}
    \centering
    \caption{Data and classifying of segregation types for the experiments performed, with the final row indicating a prediction for an industrial-scale silo.}
    \begin{tabular}{ccccccccccc}  
        Symbol & Particles & Fluid & \makecell{$d_s$\\(\textmu m)} & \makecell{$d_b$\\(\textmu m)} & \makecell{$H$\\(cm)} & \makecell{$A_r$} & \makecell{$D$\\(mm)} & $St$ & $Re_p$ & \makecell{Seg\\type}\\ \hline
        \raisebox{0.8ex}{\custommarker{circle}} & Al & Air & 64 & 150 & 6 & 1.5 & 6 & 10.4 & 43.6 & A\\
        \raisebox{0.8ex}{\custommarker{smallcircle}} & Al & Air & 64 & 150 & 6 & 1.5 & 4 & 10.4 & 43.6 & A\\
        \raisebox{0.8ex}{\custommarker{circle}} & Al & Air & 64 & 150 & 6 & 0.75 & 6 & 10.4 & 43.6 & A\\
        \raisebox{0.8ex}{\custommarker{circle}} & Al & Air & 64 & 150 & 6 & 0.38 & 6 & 10.4 & 43.6 & A\\
        \raisebox{0.8ex}{\custommarker{circle}} & Al & Air & 64 & 150 & 12 & 1.5 & 6 & 5.2 & 43.6 & A\\
        \raisebox{0.8ex}{\custommarker{smalldiamond}} & Al & Propanol & 64 & 150 & 6 & 1.5 & 4 & $1.7 \mathrm{e}{-4}$ & 0.35 & B/C\\
        \raisebox{0.8ex}{\custommarker{diamond}} & Al & Propanol & 64 & 150 & 6 & 1.5 & 6 & $1.7 \mathrm{e}{-4}$ & 0.35 & C\\
        \raisebox{0.8ex}{\custommarker{orangediamond}} & Al & Propanol & 64 & 150 & 6 & 0.75 & 6 & $1.7 \mathrm{e}{-4}$ & 0.35 & B\\
        \raisebox{0.8ex}{\custommarker{orangediamond}} & Al & Propanol & 64 & 150 & 6 & 0.38 & 6 & $1.7 \mathrm{e}{-4}$ & 0.35 & B\\
        \raisebox{0.8ex}{\custommarker{smallsquare}} & Al & Oil & 64 & 150 & 6 & 1.5 & 4 & $3.2 \mathrm{e}{-7}$ & $8 \mathrm{e}{-4}$ & B/C\\
        \raisebox{0.8ex}{\custommarker{greensquare}} & Al & Oil & 64 & 150 & 6 & 1.5 & 6 & $3.2 \mathrm{e}{-7}$ & $8 \mathrm{e}{-4}$ & C\\
        \raisebox{0.8ex}{\custommarker{smallsquare}} & Al & Oil & 64 & 150 & 6 & 1 & 4 & $3.2 \mathrm{e}{-7}$ & $8 \mathrm{e}{-4}$ & B/C\\
        \raisebox{0.8ex}{\custommarker{greensquare}} & Al & Oil & 64 & 150 & 6 & 1 & 6 & $3.2 \mathrm{e}{-7}$ & $8 \mathrm{e}{-4}$ & C\\
        \raisebox{0.8ex}{\custommarker{bigsquare}} & Al & Oil & 64 & 150 & 6 & 1 & 8 & $3.2 \mathrm{e}{-7}$ & $8 \mathrm{e}{-4}$ & C\\
        \raisebox{0.8ex}{\custommarker{orangesquare}} & Al & Oil & 64 & 150 & 6 & 0.75 & 6 & $3.2 \mathrm{e}{-7}$ & $8 \mathrm{e}{-4}$ & B\\
        \raisebox{0.8ex}{\custommarker{greensquare}} & Al & Oil & 64 & 150 & 12 & 1.5 & 6 & $1.6 \mathrm{e}{-7}$ & $8 \mathrm{e}{-4}$ & B/C\\
        \raisebox{0.8ex}{\custommarker{orangesquare}} & Al & Oil & 64 & 150 & 6 & 0.38 & 6 & $3.2 \mathrm{e}{-7}$ & $8 \mathrm{e}{-4}$ & B\\
        \raisebox{0.8ex}{\custommarker{greensquare}} & Al & Oil & 64 & 150 & 40 & 10 & 4 & $3.2 \mathrm{e}{-8}$ & $8 \mathrm{e}{-4}$ & B/C\\
        \raisebox{0.8ex}{\custommarker{triangle}} & GB1 & Air & 120 & 362 & 6 & 0.75 & 6 & 106 & 117 & A\\
        \raisebox{0.8ex}{\custommarker{orangetriangle}} & GB1 & Oil & 120 & 362 & 6 & 0.75 & 6 & $2.6 \mathrm{e}{-6}$ & $4 \mathrm{e}{-3}$ & B\\
        \raisebox{0.8ex}{\custommarker{blackstar}} & GB2 & Air & 362 & 1090 & 33 & 3.3 & 10 & 1592.4 & 612 & A\\
        \raisebox{0.8ex}{\custommarker{blackstar}} & GB2 & Air & 362 & 1090 & 19 & 1.3 & 14 & 2766 & 612 & A\\
        \raisebox{0.8ex}{\custommarker{blackstar}} & GB2 & Air & 362 & 1090 & 19 & 0.68 & 16 & 2766 & 612 & A\\
        \raisebox{0.8ex}{\custommarker{greenstar}} & GB2 & Soapy water & 362 & 1090 & 33 & 3.3 & 10 & 0.095 & 150 & C\\
        \raisebox{0.8ex}{\custommarker{greenstar}} & GB2 & Soapy water & 362 & 1090 & 19 & 1.3 & 14 & 0.16 & 150 & C\\ 
        \raisebox{0.8ex}{\custommarker{orangestar}} & GB2 & Soapy water & 362 & 1090 & 19 & 0.68 & 16 & 0.16 & 150 & B\\  
        \hline
        \raisebox{0.8ex}{\custommarker{invertedblue}} & Al & Air & 64 & 150 & 5000 & 1.25 & - & 0.012 & 43.6 & C \\
    \end{tabular}
    \label{tab:all}
\end{table*}

\begin{figure}
    \centering
    \includegraphics{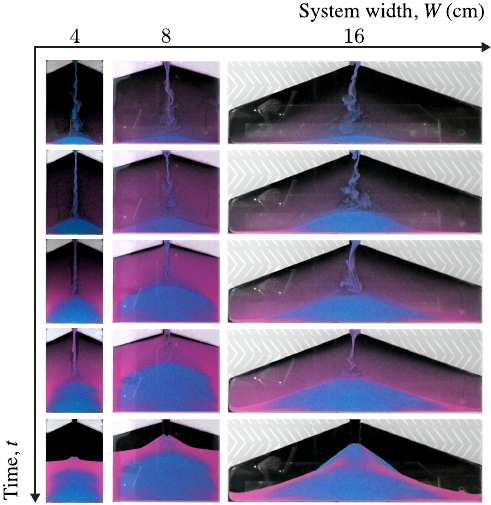}
    \caption{Flow and segregation evolution of alumina particles of two sizes in propanol for different system widths $W$ = 4, 8 and 16 cm, and fixed height $H=6$~cm. Time increases from top to bottom. Large particles are blue and small particles are pink.
    }
    \label{fig:time_series}
\end{figure}

Notice that the selected experimental parameters in Table \ref{tab:all} covers a wide range of Stokes and Reynolds numbers, over 12 and 6 orders of magnitude, respectively. The slenderness varies over only one order, which has been chosen to cover a range of geometries used in practice.

In the following it will be demonstrated that only $A_r$ and $St$ are actually required for delineating the different segregation patterns, and the upscaling of these patterns from laboratory to large-scale silos. In particular, the slenderness is clearly important when considering the extent of lateral motion in highly slender ($A_r\gg 1$) or wide ($A_r\rightarrow 0$) silos. The Stokes number represents the response of the particles to fluid flow, as in whether the particles follow the fluid or not, and should thus be considered. Indeed, the Stokes number has been shown to capture the transition from inertia-dominated dynamics ($\mathrm{St} \gg 1$) to viscous-dominated dynamics ($\mathrm{St} \ll 1$), and has been used to describe run-out behavior in granular collapse~\cite{topin2012collapse,bougouin2018granular,courrech2003granular,rauter2021compressible,yang2021size,tang2024granular}, segregation in shear flows \cite{cui2020generalized,zhou2020particle,cui2021viscous,cui2022particle,ferdowsi2017river,gonzalez2023bidisperse,gonzalez2024forces} and during sedimentation ~\cite{weiland1984instabilities,abbas2006dynamics,deboeuf2011segregation,zhang2025hindered}. 
On the other hand, the Reynolds number describes the nature of the fluid flow, whether it is laminar or turbulent. Therefore, it may not be expected to be pivotal for delineating the transitions between the various segregation patterns. Indeed, this point will be demonstrated using the experiments.

\begin{figure*}
    \centering
    \includegraphics[width=\textwidth]{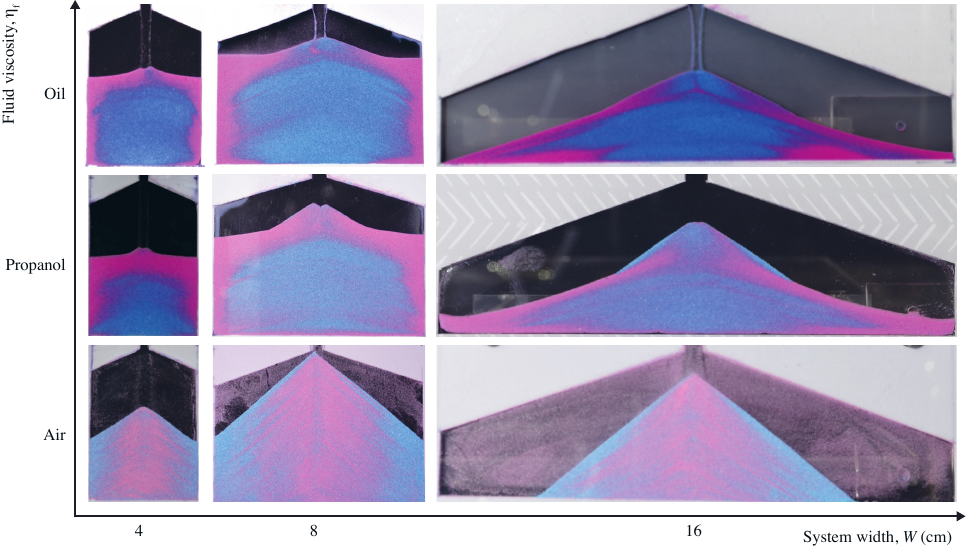}
    \caption{Segregation patterns of alumina particles with systems of different width $W$, and varying fluid. Large particles are blue and small particles pink. For all cases, $H=6$~cm, $W/D \geqslant 10$ and $T=0.2$ cm.}
    \label{fig:Alumina_figs}
\end{figure*} 

\begin{figure}
\centering
    \includegraphics[width=0.9\columnwidth]{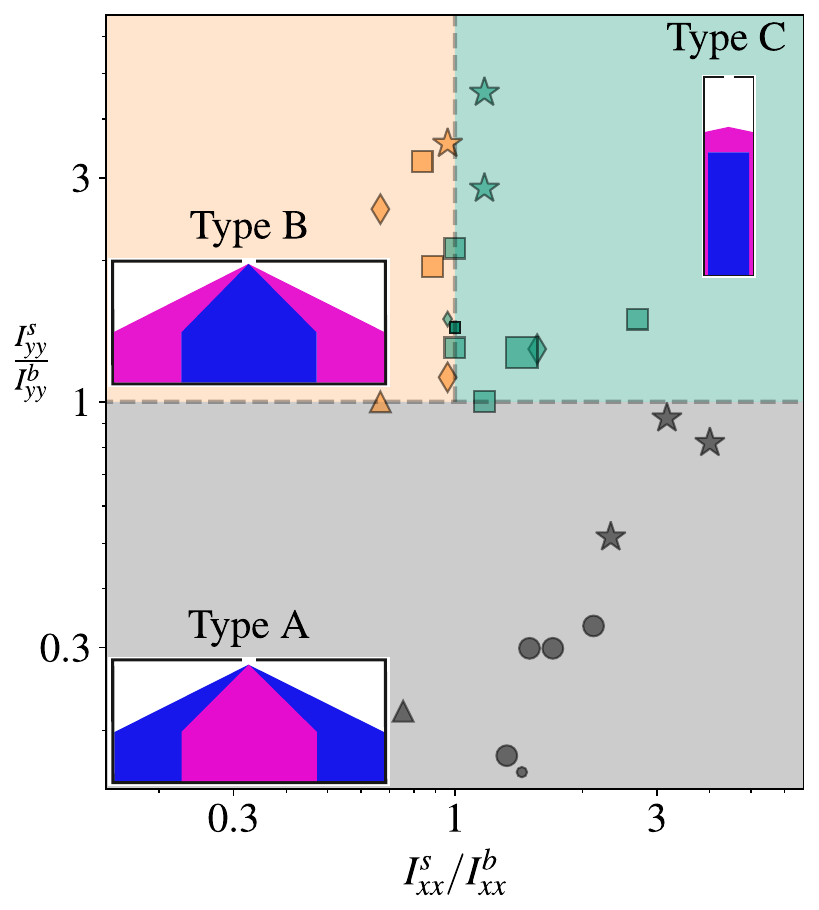}
    \caption{Classifying segregation patterns based on image analysis. Phase diagram in terms of the normalised second moments of area around the $x$ and $y$ axes, $I_{yy}^{s} / I_{yy}^{b}$ and $I_{xx}^{s} / I_{xx}^{b}$, respectively. Definitions of each marker are shown in Table \ref{tab:all}.}
    \label{fig:Img_analysis}
\end{figure} 

\begin{figure}
\centering
    \includegraphics[width=\columnwidth]{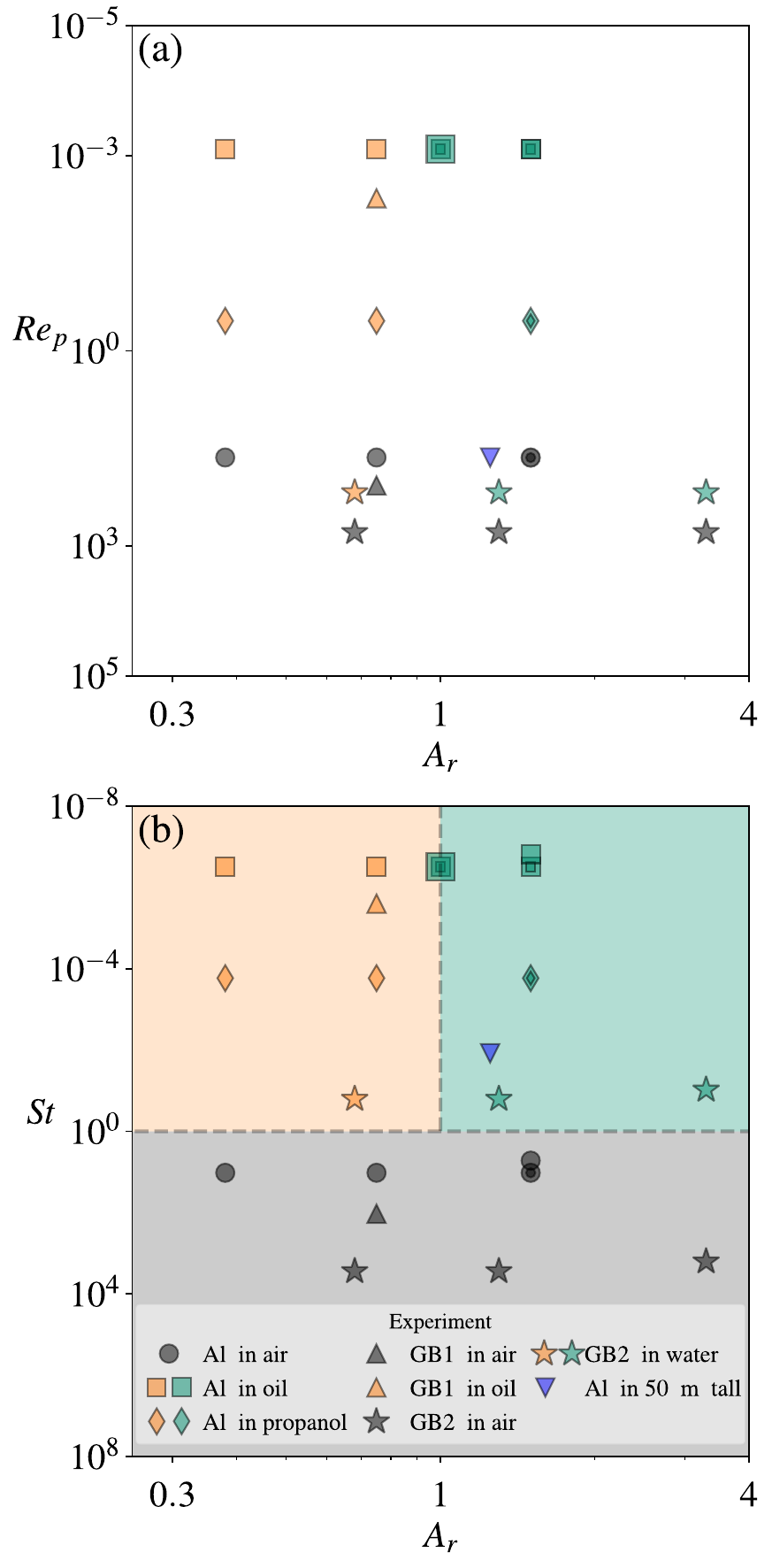}
    
    \caption{Classifying segregation patterns based on dimensionless numbers for the experiments listed in Table~\ref{tab:all}, and for a hypothetical industrial silo indicated by the blue marker, also detailed in Table~\ref{tab:all}. (a) Reynolds number versus slenderness. (b) Stokes number versus slenderness. Markers indicate experiments according to Table \ref{tab:all}.}
    \label{fig:Re_analysis}
\end{figure}

After the cessation of the granular flow, an image of each static heap was captured using a camera, as shown for example in Figure \ref{fig:lab_model}(b). A color threshold was then used on the image to separate regions of large (Figure \ref{fig:lab_model}(c)) and small (Figure \ref{fig:lab_model}(d)) grains. The union of the two segmented regions recovers the entire heap. Following segmentation, the second moments of area along the y- and x-axes are computed for each grain type as follows:  

\begin{equation}
I_{xx}^{\alpha}=\int_{A_{\alpha}} y^2 \, da,\quad I_{yy}^{\alpha}=\int_{A_{\alpha}} x^2 \, da,
\end{equation}
where $\alpha \in \{s,b \}$ indicate the small ($s$) and big ($b$) particles, respectively, and $A$ indicates the area corresponding to a considered grain type. The normalized values are indicated on the respective plots in Figures \ref{fig:lab_model}(c) and (d). These quantities, which are calculated about the centroid of the entire heap, characterize the amount of spatial spreading of a considered grain type. 

The temporal evolution of alumina particles immersed in propanol are illustrated both using movies of silo filling in oil and propanol in the Supplementary Information, as well as through snapshots for three different silo widths in Figure~\ref{fig:time_series}, where the height is fixed. The figure shows the progression of the pattern, with the stream of particles first hitting the base and gradually forming a deposit. As time increases, the large blue particles tend to settle near the center, with the small pink particles carried outward by the fluid. Observation of video recordings indicates that the flow of particles and fluid is redirected when it reaches the pile, from an initially gravity-driven downwards flow to a horizontally outwards flow towards the walls. During this outwards flow, larger particles settle more quickly than smaller ones, and as a result we have more small particles near the system walls. 
Note that as the experiment occurs inside a closed cell with two chambers, any flux of material and fluid into the bottom chamber must be balanced by a similar flux upwards. The interaction that develops between the two chambers can influence the settling dynamics in the lower chamber, which is studied here. We assume that the impact of this on the final pattern is negligible. 



The final patterns after the deposition of Alumina particles in three different fluids (air, propanol and oil) for three different system widths are shown in Figure~\ref{fig:Alumina_figs}. These images clearly show a range of patterns. In all cases involving air, small particles tend to accumulate near the center of the heap, while large particles are found closer to the silo walls. In contrast, in propanol and oil, the small particles are always at the top, but the deposit appears as either a sandpile or a column with large particles predominantly settling near the center. To distinguish these segregation patterns, we employ image analysis by calculating the second moments of area for each particle type about the centroid of the entire heap. This metric captures both the extent of spreading and the spatial positioning of each particle type relative to the overall heap. The ratios between the moment of inertia of the areas of the small and big particles ($I_{xx}^s/I_{xx}^b$ and $I_{yy}^s/I_{yy}^b$) along the $y$ and $x$ axes are characteristics of the observed patterns. We distinguish three patterns (namely type A, B and C) through transitions below and above unity of these ratios, as shown in Figure~\ref{fig:Img_analysis}, with the corresponding patterns schematised in insets. The last column in Table \ref{tab:all} lists the resulting segregation type for each of the experiments. In all cases involving air, we find a type A pattern, where as in propanol and oil, we find either type B or C.
A similar analysis is carried out for two kind of glass beads (GB1 and GB2), with corresponding classification also in Table \ref{tab:all}.  

Next, we inspect whether a single phase diagram in terms of the physical dimensionless numbers (Eqs. \ref{eq:aspect_ratio}-\ref{eq:rep}) could demarcate the three patterns. An attempt to classify the patterns using the Reynolds number ($Re_p$) fails to clearly distinguish types A, B and C patterns (see Figure~\ref{fig:Re_analysis}(a)). On the other hands, using  the Stokes number ($St$) in Figure~\ref{fig:Re_analysis}(b) achieves the purpose: the three different pattern types appear in regions below and above one in the corresponding dimensionless numbers. Specifically, while transitioning from viscous ($St < 1$) to inertial ($St > 1$) regimes, small particles are found to be deposited on the sides or at the top of the silo, meaning type B or C.

The analysis in this paper demonstrates how the Stokes number $St$ and the slenderness of the silo $A_r$ could be used to predict the type of segregation pattern emerging during fluid-submerged silo filling. Additionally, and perhaps more importantly, using these two dimensionless numbers and the obtained phase diagram could be adopted as an upscaling framework from laboratory- to industry-scale silos. For instance, according to Figure~\ref{fig:Re_analysis}(b), a 50m tall and 40m wide silo filled with air and with a similar bi-mixture of grain sizes as studied here is expected to yield type C segregation patterns. This is consistent with field observations reported in \cite{engblom2012effects,engblom2012segregation,zigan2008theoretical}. The results thus underscore the necessity of considering the role of air and incorporating the Stokes number in segregation models, especially for large-scale systems, and motivate the development of new scaling relations for continuum descriptions. While the current study systematically probed the role of key dimensionless numbers in 2D setups, a complementary investigation involving 3D setups would help to extend its generality. In addition, the role of discharge rate on the segregation dynamics remains an open and important question.

\bibliographystyle{apsrev4-1}
\bibliography{bib}

\appendix
\section{Coloring method}\label{app:colouring}
In order to color the alumina grains, we used normal fabric dye mixed with water. After $24 - 48$ hours of coloring the grains, we oven-dried the grains at $105^{\circ}C$ for $24$ hours. 
The glass beads were dyed by first applying a coat of ESP aerosol on the surface of the glass beads, then allowed to dry for $2$ hours. We then applied a coat of acrylic enamel paint over the dried glass beads. After air drying for 2 hours and then further oven drying at $105^{\circ}C$ for $24$ hours, the glass beads were sieved to obtain the respective size fractions and used in the experiments.

\end{document}